\documentclass[english]{article}
\usepackage[T1]{fontenc}
\usepackage[latin9]{inputenc}
\usepackage{float}
\usepackage{amsmath}
\usepackage{stackrel}
\usepackage{graphicx}

\makeatletter
\usepackage{amsfonts,amssymb}
\usepackage{latexsym}
\usepackage{epsfig}
\usepackage{cite}
\usepackage{graphicx}
\usepackage[colorlinks,linkcolor=blue]{hyperref}
\newcommand{\be}{\begin{equation}}
\newcommand{\ee}{\end{equation}}

\makeatother

\usepackage{babel}
\begin{document}
{}~ \hfill\vbox{\hbox{CTP-SCU/2019014}}\break
\vskip 3.0cm
\centerline{\Large \bf Construct  $\alpha^{\prime}$ corrected or loop corrected solutions without curvature singularities}

\vspace*{10.0ex}
\centerline{\large Peng Wang, Houwen Wu, Haitang Yang and Shuxuan Ying}
\vspace*{7.0ex}
\vspace*{4.0ex}
\centerline{\large \it College of Physics}
\centerline{\large \it Sichuan University}
\centerline{\large \it Chengdu, 610065, China} \vspace*{1.0ex}
\vspace*{4.0ex}

\centerline{pengw@scu.edu.cn, iverwu@scu.edu.cn, hyanga@scu.edu.cn, ysxuan@stu.scu.edu.cn}
\vspace*{10.0ex}
\centerline{\bf Abstract} \bigskip \smallskip

For the bosonic gravi-dilaton system,   we provide systematical approaches to construct non-perturbative string cosmological solutions without curvature singularities, which can match the perturbative solution  to any order in $\alpha'$ expansion. When higher order  $\alpha'$ corrections are calculated, they can be straightforwardly plugged in to generate compatible non-perturbative evolutions without curvature singularities.  We also give a (phenomenological) map between   $\alpha^{\prime}$ corrected EOM and loop corrected EOM. This map enables us to easily generate a loop corrected solution from an $\alpha^{\prime}$ corrected solution, and vice versa, therefore substantially enlarges the solution space.

\vfill
\eject
\baselineskip=16pt
\vspace*{10.0ex}

\section{Introduction}

An important challenge for string theory is to show how the big-bang
singularity could be resolved. In the Einstein gravity, the big-bang
singularity is the initial singularity. Nevertheless, in the traditional
(tree level) string cosmology, a ``scale-factor'' duality emerges
\cite{Tseytlin:1991wr,Veneziano:1991ek,Meissner:1991zj, Sen:1991zi,Sen:1991cn,Tseytlin:1991xk}.
This duality combined with time reversal yields a new phase: the pre-big-bang \cite{Veneziano:2000pz,Gasperini:2002bn,Gasperini:2007vw,Gasperini:1992em}.
The big-bang singularity splits the pre-big-bang and post-big-bang
into two disconnected regions. To be specific, we set the spacetime
dimensionality to be $D=d+1$ and  work with bosonic string theory. The scale-factor duality turns out
to be a special case of a more general symmetry, the $O(d,d)$ symmetry.
This duality has no descendant in the Einstein gravity since the dilaton
transform nontrivially.

Beyond the perturbative regime, the tree level\footnote{If not specified, ``tree level'' indicates the lowest order in both
$\alpha'$ and loop.} string effective action receives two kinds of corrections: the higher-derivative
expansion, controlled by the squared string length $\alpha'$, and
the higher-genus expansion, controlled by the string coupling $g_{s}=e^{2\phi}$.
Ignoring matter sources, the most general perturbative form of the
string effective action has the following structure

\begin{eqnarray}
I & = & \int\:d^{d+1}x\sqrt{-g}\bigg\{ e^{-2\phi}\Big[(R+4(\partial\phi)^{2}-
\frac{1}{12}{\cal H}^{2})+\frac{\text{\ensuremath{\alpha'}}}{4}(R_{\mu\nu\sigma\rho}R^{\mu\nu\sigma\rho}+\cdots)+{\cal O}(\alpha'^{2})\big]\nonumber\\
 & + & \Big[(c_{R}^{1}R+c_{\phi}^{1}(\partial\phi)^{2}+c_{{\cal H}}^{1} {\cal H}^{2})+\alpha'(c_{\alpha'R}^{1}R_{\mu\nu\sigma\rho}R^{\mu\nu\sigma\rho}+\cdots)+{\cal O}(\alpha'^{2})\Big]\nonumber\\
 & + & e^{2\phi}\Big[(c_{R}^{2}R+c_{\phi}^{2}(\partial\phi)^{2}+c_{{\cal H}}^{2} {\cal H}^{2})+\alpha'(c_{\alpha'R}^{2}R_{\mu\nu\sigma\rho}R^{\mu\nu\sigma\rho}+\cdots)+{\cal O}(\alpha'^{2})\Big]\nonumber\\
 & + & \cdots\bigg\},
\label{eq:complete action}
\end{eqnarray}
where $\phi$ is the dilaton and ${\cal H}_{\mu\nu\rho}=3\partial_{[\mu}b_{\nu\rho]}$
is the field strength of the antisymmetric Kalb-Ramond field $b_{\mu\nu}$.
For simplicity, we set $b_{\mu\nu}=0$ in this paper. All the coefficients
$c_{[\cdots]}^{i}$ are yet unknown. Each line contains a full expansion
in $\alpha'$. In terms of genus, the first line is the tree level
terms with complete stringy contributions, the second line is the
full one-loop contribution, and so on. Throughout this paper, we always working with   FLRW background

\begin{equation}
ds^2= - dt^2 +a^2(t) \delta_{ij} dx^i dx^j,
\label{FLRW}
\end{equation}
with the Hubble parameter $H\equiv \dot a/a$. The traditional tree level
cosmology does not take into account the $\alpha'$ and loop corrections,
thus is valid only in the perturbative regime $g_{s}\to0$ and $\alpha' H^2\to0$.
As the universe approaches the big-bang region, there would be $g_{s}\to1$,
$\alpha' H^2\to1$ or both. It is then natural to anticipate the $\alpha'$
or loop corrections could regularize the big-bang singularity. Indeed,
by implementing some non-local dilaton potentials which account for
non-perturbative effects caused by the dilaton, the loop corrections
could smooth out the singularity \cite{Gasperini:1992em,Gasperini:2003pb,Gasperini:2004ss, Gasperini:book}.

However, there is not much progress on how to resolve the big-bang
singularity with $\alpha'$ corrections. The main reason is that,
the higher-derivative $\alpha'$ corrections usually would change
the order of the differential equations in the equations of motion
(EOM). At the tree level, the EOM are second order differential equations,
at the first order in $\alpha'$, the EOM become fourth order differential
equations, and so on.
In \cite{Easson:2003ia}, by assuming the heterotic
string admits non-singular constant curvature solutions in the Einstein
frame, an $O(d,d)$ violating first order $\alpha'$ correction was
chosen. In terms of the scale factor $a(t)$, the EOM, as expected,
are fourth order differential equations. A carefully designed effective
dilaton potential was further brought in to support a non-singular
evolution from an early-time de-Sitter phase to a late-time Minkowski
spacetime.

Notwithstanding little hope to conduct analysis on the
higher order $\alpha'$ corrections, inspiringly, for the first order
$\alpha'$ correction, by using some field redefinitions, it turns
out that the fourth order derivatives can be eliminated \cite{Zwiebach:1985uq}.
Thus the EOM are still second order differential equations. This nice
property enables the authors of ref. \cite{Gasperini:1996fu} to numerically verify that the perturbative string vacuum could connect with some ``fixed-points'', at the cost of  the scale-factor duality.


Recently, the situation has changed by the remarkable work of Hohm and
Zwiebach \cite{Hohm:2015doa,Hohm:2019ccp,Hohm:2019jgu}. Early works
in refs. \cite{Veneziano:1991ek, Sen:1991zi,Sen:1991cn, Meissner:1991zj}
showed that for cosmological background, all orders in $\alpha'$
expansion possess an $O(d,d)$ symmetry. Moreover, to the first order
in $\alpha'$, the $O(d,d)$ matrix can maintain the standard form
in term of $\alpha'$ corrected fields \cite{Meissner:1996sa}. With a reasonable assumption
that this property also holds for all orders in $\alpha'$, Hohm and
Zwiebach proved that for time dependent configurations, only
first order time derivatives of the fields appear in the action. An
immediate consequence of this striking simplification is that the
EOM with complete $\alpha'$ corrections are still second order differential equations.

The Hohm-Zwiebach action paves the way to seriously address the non-perturbative
features sourced by $\alpha'$ corrections. This remarkable
result leads them to show that, in bosonic string theory, non-perturbative
de-Sitter (dS) vacua are admitted by including complete $\alpha'$
corrections. In \cite{Krishnan:2019mkv}, the analogy in the Einstein
frame is then discussed. In our recent work \cite{Wang:2019mwi},
we showed that similar stories occur for configurations depending
on a single space coordinate, and non-perturbative Anti-de-Sitter
(AdS) vacua are also allowed. Furthermore, we   conjectured
that the non-perturbative AdS and dS vacua might not be able to coexist
in bosonic string theory.

The Hohm-Zwiebach action also sheds light on the resolution of  the big-bang
singularity. However, straightforward perturbative calculation does
not work. As shown in ref. \cite{Hohm:2019jgu}, the Hubble
parameter and $O(d,d)$ invariant dilaton  calculated order by order
in $\alpha'$ are

\begin{eqnarray}
H(t) & = & \frac{1}{\sqrt{d}t}-\frac{5}{4}\frac{1}{d^{3/2}}\frac{\alpha'}{t^{3}} + h_2 \frac{\alpha'^{2}}{t^{5}} + h_3 \frac{\alpha'^{3}}{t^{7}} +\cdots, \nonumber\\
\Phi(t) & = &-\frac{1}{2}\log\left(\gamma^{2}t^{2}\right)-\frac{1}{2}\frac{t_{0}^{2}}{t^{2}}+\omega_{2}\frac{t_{0}^{4}}{t^{4}}+\omega_{3}\frac{t_{0}^{6}}{t^{6}}+\cdots,
\label{eq:HZ perturbative solution}
\end{eqnarray}
where the coefficients $h_{i}$ and $\omega_{i}$ are yet undetermined, $\gamma$ is an integration constant and $t_{0}\equiv\frac{\sqrt{\alpha'}}{\sqrt{2d}}$. The $O(d,d)$ invariant dilaton $\Phi$ is defined as $e^{-\Phi}=\sqrt{-g} e^{-2\phi}$.
Obviously, higher order terms are more and more singular. After realizing the above solution
is actually valid in the perturbative regime $t\to\infty$  ($\alpha'\to0$), in \cite{Wang:2019kez},
we constructed  non-perturbative solutions which are non-singular (non-singular in this paper refers to the curvature, but not to the string coupling behaviour\footnote{We wish to address that the term ``non-singular'' in our previous work \cite{Wang:2019kez} and this paper means that
the curvature and $O(d,d)$ dilaton $\Phi$ have no singularities. However, the string coupling which is controlled by the physical dilaton $g_s = e^{2\phi}= \sqrt{-g} e^\Phi$,  blows up as $t\to\infty$ in our solutions. So,  more precisely, what we provide are solutions without curvature singularities.   We are indebted to the anonymous referee to help us clarify  this confusion. In an upcoming paper, we will demonstrate that the string coupling can also be regularized by introducing a non-trivial Kalb-Ramond field $B(t)$ into the solutions \cite{Guo}.}) in the whole regime $t\in(-\infty,\infty)$ for nonvanishing $\alpha'$.
The term ``non-perturbative'' refers to that the domain of the solution
covers the non-perturbative regime and all $\alpha'$ corrections
are included. Those non-singular non-perturbative solutions are justified
by matching the first two orders of the perturbative solution eq. (\ref{eq:HZ perturbative solution}) exactly and having the same expansion behaviors at higher orders, in the perturbative regime $t\to\infty$ ($\alpha'\to0$).

Hitherto, in the perturbative $\alpha'$  expansion,   orders  higher than one are unknown.
So the solutions constructed in \cite{Wang:2019kez} only need to the match the first two orders. An inspiring question naturally arises: Is there any guidance to construct  non-singular non-perturbative solutions when  orders higher than one are calculated in the future? The trial and error is a very inefficient method and becomes unpractical for high orders. One of the purposes of this paper is to do this job. We are going to provide two formulas  to easily construct ``more accurate'' non-singular non-perturbative solutions, which can match the perturbative results to an arbitrary order. In contrast to the perturbative solution which is more and more singular at higher orders, every term in our solutions is non-singular. One more $\alpha'$ correction is provided, one more non-singular term is fixed. This process continues to any order.

Moreover, we find a very useful and suggestive (phenomenological) map between the EOM corrected by $\alpha'$ and the EOM corrected by loops.  The effective dilaton potentials which represent loop corrections can be mapped to  some functions of  $\alpha'$ corrections.  With this map, one can easily generate an $\alpha'$ corrected solution from a loop corrected solution, or vice versa.  It turns out it is much easier to construct $\alpha'$ corrected solutions with our method than to find loop corrected solutions. Therefore, this map  substantially enlarges the solution space of the traditional string cosmology, and one may analysis more scenarios. Furthermore, the new loop corrected solutions generated from the $\alpha'$ corrected solutions we constructed are more consistent and reasonable than those given in literature.

The reminder of this paper is outlined as follows. In section 2, we show how to construct non-singular non-perturbative solutions to any order in $\alpha'$ expansion.  In section 3, we present a (phenomenological) map between  $\alpha'$ corrected EOM and loop corrected EOM. We also give some examples.   Section 4 is the conclusion.

\section{Solutions without curvature singularities to an arbitrary order in $\alpha'$ }

It is well known that, for FLRW background (\ref{FLRW}), the tree level string
effective action can be put into an explicit $O(d,d)$ covariant form.
This is also true at the first order in $\alpha'$ with appropriate
field redefinitions \cite{Meissner:1996sa}. Based on a reasonable assumption that, to all
orders in $\alpha'$, the standard $O(d,d)$ matrix can be maintained
by field redefinitions, in \cite{Hohm:2019ccp,Hohm:2019jgu}, a substantial
simplification on the $\alpha'$ corrections is achieved:

\begin{eqnarray}
I & = & \int d^{D}x\sqrt{-g}e^{-2\phi}\left(R+4\left(\partial\phi\right)^{2}+\frac{1}{4}\alpha^{\prime}\left(R^{\mu\nu\rho\sigma}R_{\mu\nu\rho\sigma}+\ldots\right)+\alpha^{\prime2}(\ldots)+\ldots\right),\label{eq:original action with alpha}\\
 & = & \int dte^{-\Phi}\left(-\dot{\Phi}^{2}+\sum_{k=1}^{\infty}\left(\alpha^{\prime}\right)^{k-1}c_{k} \mathrm{tr}\left(\dot{\mathcal{S}}^{2k}\right)\right),\label{eq:Odd action with alpha}
\end{eqnarray}

\noindent where $\Phi\left(t\right)=2\phi\left(t\right)-\log\sqrt{-g}$
is the $O\left(d,d\right)$ invariant dilaton. Kalb-Ramond field is
set to be zero for simplicity. The first line is the classical action in a general
background. The second line is the Hohm-Zwiebach action in FLRW metric.
The $2d\times2d$ standard $O(d,d)$ matrix $\mathcal{S}$ is defined
as

\noindent
\begin{equation}
\mathcal{S}=\left(\begin{array}{cc}
0 & a^{2}\left(t\right)\\
a^{-2}\left(t\right) & 0
\end{array}\right).
\end{equation}

\noindent Thus far, in the Hohm-Zwiebach action (\ref{eq:Odd action with alpha}),
only $c_{1}=-\frac{1}{8}$ and $c_{2}=\frac{1}{64}$ for the bosonic
string theory ($c_{2}=\frac{1}{128}$ for heterotic string and $c_{2}=0$
for type II strings) are calculated through the beta functions of
the non-linear sigma model, and $c_{k\geq3}$ are undetermined constants.
The EOM (generalized Friedmann equations) of (\ref{eq:Odd action with alpha}) are given by

\begin{eqnarray}
\ddot{\Phi}+\frac{1}{2}Hf\left(H\right) & = & 0,\nonumber \\
\dot{\Phi}^{2}+g\left(H\right) & = & 0,\nonumber \\
\frac{d}{dt}\left(e^{-\Phi}f\left(H\right)\right) & = & 0,\label{eq:Original EoM}
\end{eqnarray}

\noindent with

\begin{eqnarray}
H\left(t\right) & = & \frac{\dot{a}\left(t\right)}{a\left(t\right)},\nonumber \\
f\left(H\right) & = & d\sum_{k=1}^{\infty}\left(-\alpha^{\prime}\right)^{k-1}2^{2\left(k+1\right)}kc_{k}H^{2k-1} =-2dH-2d\alpha^{\prime}H^{3}+\mathcal{O}\left(\alpha^{\prime2}\right),\nonumber \\
g\left(H\right) & = & d\sum_{k=1}^{\infty}\left(-\alpha^{\prime}\right)^{k-1}
2^{2k+1}\left(2k-1\right)c_{k}H^{2k}=-dH^{2}-\frac{3}{2}
d\alpha^{\prime}H^{4}+\mathcal{O}\left(\alpha^{\prime2}\right),\label{eq:EOM fh gh}
\end{eqnarray}

\noindent where $H\left(t\right)$ is the Hubble parameter. Note

\begin{equation}
g'(H)=Hf'(H),\quad{\rm and}\quad g(H)=Hf(H)-\int_{0}^{H}f(x)dx,
\label{eq:gf relation}
\end{equation}
where $f'(H)\equiv \frac{d}{dH} f(H)$. The Hohm-Zwiebach action can be recast as
\begin{equation}
I_{HZ}=\int dte^{-\Phi}\left(-\dot{\Phi}^{2}+g(H)-Hf(H)\right).
\end{equation}
In the perturbative regime  $|t|\to\infty$  ($\alpha'\to 0$), using (\ref{eq:EOM fh gh}),  the EOM can be solved iteratively to arbitrary order in $\frac{\sqrt{\alpha'}}{t}$,

\begin{eqnarray}
H\left(t\right) & = & \frac{\sqrt{2}}{\sqrt{\alpha^{\prime}}}\left[\frac{t_{0}}{t}-160c_{2}\frac{t_{0}^{3}}{t^{3}}+\frac{256\left(770c_{2}^{2}+19c_{3}\right)}{3}\frac{t_{0}^{5}}{t^{5}}\right.\nonumber\\
 &  & \left.-\frac{2048\left(88232c_{2}^{3}+4644c_{3}c_{2}+41c_{4}\right)}{5}\frac{t_{0}^{7}}{t^{7}}+\mathcal{O}\left(\frac{t_{0}^{9}}{t^{9}}\right)\right], \quad t_{0}\equiv\frac{\sqrt{\alpha'}}{\sqrt{2d}} \nonumber\\
\Phi\left(t\right) & = & -\frac{1}{2}\log\left(\beta^2\frac{t^{2}}{t_{0}^{2}}\right)-32c_{2}\frac{t_{0}^{2}}{t^{2}}+\frac{256\left(44c_{2}^{2}+c_{3}\right)}{3}\frac{t_{0}^{4}}{t^{4}} \nonumber\\
 &  & -\frac{2048\left(6976c_{2}^{3}+352c_{3}c_{2}+3c_{4}\right)}{15}\frac{t_{0}^{6}}{t^{6}}+\mathcal{O}\left(\frac{t_{0}^{8}}{t^{8}}\right),
\label{eq:perturbative solution}
\end{eqnarray}
and
\begin{eqnarray*}
f\left(H\left(t\right)\right) & = & -2dH-128c_{2}d\alpha^{\prime}H^{3}+768c_{3}d\alpha^{\prime2}H^{5}-4096c_{4}d \alpha^{\prime3}H^{7}+\mathcal{O}\left(\alpha^{\prime4}H^{9}\right), \nonumber\\
 & = & \frac{\sqrt{d}}{t_{0}}\left[-\frac{2t_{0}}{t}+64c_{2}\frac{t_{0}^{3}}{t^{3}}-\frac{512\left(50c_{2}^{2}+c_{3}\right)}{3}\frac{t_{0}^{5}}{t^{5}}\right. \nonumber\\
 &  & \left.+\frac{4096\left(2632c_{2}^{3}+124c_{3}c_{2}+c_{4}\right)}{5} \frac{t_{0}^{7}}{t^{7}}+\mathcal{O}\left(\frac{t_{0}^{9}}{t^{9}}\right)\right],\nonumber\\
g\left(H\left(t\right)\right) & = & -dH^{2}-96c_{2}d\alpha^{\prime}H^{4} +640c_{3}d\alpha^{\prime2}H^{6}-3584c_{4}d\alpha^{\prime3} H^{8}+\mathcal{O}\left(\alpha^{\prime4}H^{10}\right),\nonumber\\
 & = & \frac{1}{t_{0}^{2}}\left[-\frac{t_{0}^{2}}{t^{2}}+128c_{2}\frac{t_{0}^{4}}{t^{4}}-\frac{2048\left(50c_{2}^{2}+c_{3}\right)}{3}\frac{t_{0}^{6}}{t^{6}}\right.\nonumber\\
 &  & \left.+\frac{8192\left(24448c_{2}^{3}+1136c_{3}c_{2}+9c_{4}\right)}{15}\frac{t_{0}^{8}}{t^{8}}+\mathcal{O}\left(\frac{t_{0}^{10}}{t^{10}}\right)\right],
\end{eqnarray*}

\noindent where $\beta^{2}=\gamma^{2}t_{0}^{2}=\gamma^{2}\frac{\alpha^{\prime}}{2d}$ is an integration constant, $t_{0}\equiv\frac{\sqrt{\alpha'}}{\sqrt{2d}}$ and we used the universal $c_{1}=-\frac{1}{8}$.
Note for all solutions, their (scale-factor) dual solutions: $H(t)\to-H(t)$, $\Phi(t)\to\Phi(t)$,
$f(t)\to-f(t)$ and $g(t)\to g(t)$ are always implied in this paper. This solution is obviously
singular around the big-bang region $t=0$. In a recent work \cite{Wang:2019kez}
(where we set $\beta^{2}=4d$), we have constructed a pair of
non-perturbative non-singular (scale-factor) dual solutions for the
EOM (\ref{eq:Original EoM}), which exactly match the perturbative
solution (\ref{eq:perturbative solution}) in the perturbative regime,

\begin{eqnarray}
H(t) & = & -\frac{\sqrt{2}}{\sqrt{\alpha'}}\frac{\left(1-\tau^{2}\right)}{\left(1+\tau^{2}\right)^{3/2}}, \qquad \tau\equiv  \frac{t}{t_0} = \frac{\sqrt{2d}}{\sqrt{\alpha'}}t, \nonumber \\
\Phi(t) & = & -\frac{1}{2}\log\beta^2-\frac{1}{2}\log\left(1+\tau^{2}\right),\nonumber\\
f(t) & = & -\frac{2\sqrt{2}d}{\sqrt{\alpha'}}\frac{1}{\sqrt{1+\tau^{2}}} = -2d H -2d\alpha' H^3 + {\cal O}(\alpha^{\prime 2}),\nonumber \\
g(t) & = & -\frac{2d}{\alpha'}\frac{\tau^{2}}{\left(1+\tau^{2}\right)^{2}} =  -d H^2 -\frac{3}{2} d\alpha' H^4 + {\cal O}(\alpha^{\prime 2}).
\label{eq:2nd order solution}
\end{eqnarray}

After constructing the above solution which is consistent with the already known
$c_{1}$ and $c_{2}$, one may wonder when the coefficients $c_{k\ge3}$ in eq. (\ref{eq:EOM fh gh})
of higher orders are available, what the compatible non-singular solutions would be? It would be very unpleasant  if we have to do trial and error again and again. In particular, for orders very high, trial and error even becomes impossible. In the following, we are going to provide two methods to solve this problem.

Referring to the EOM (\ref{eq:Original EoM}), a very useful observation is that all other quantities could be determined by $\Phi(t)$:

\begin{eqnarray}
g\left(H\left(t\right)\right) & = & -\dot{\Phi}^{2},\nonumber \\
f\left(H\left(t\right)\right) & = & - \frac{2\sqrt{2}d}{\sqrt{\alpha'}}\beta e^\Phi,\nonumber \\
H\left(t\right) & =& \frac{\sqrt{\alpha'/2}}{\beta d}\frac{\ddot{\Phi}}{e^{\Phi}},
\label{eq:EOM in dilaton}
\end{eqnarray}
where   the integration constant has been set to be consistent with  the perturbative solution (\ref{eq:perturbative solution}).   Therefore, we only need to figure out a proper $\Phi(t)$ to make the solutions non-singular.   As $t\to\infty$ ($\alpha'\to0$), the ansatz  $\Phi(t)$ must exactly match the perturbative solution (\ref{eq:perturbative solution}), which ensures that $f(H)$ and $g(H)$ are identical to eq. (\ref{eq:EOM fh gh}). In addition, we should also check $H(t)$ is non-singular, since an inappropriate non-singular $\Phi(t)$ may lead to a singular $H(t)$.
So, the core and most difficult part is to find the right ansatz
for $\Phi(t)$. Fortunately, we  already have a successful example eq. (\ref{eq:2nd order solution}) to guide us to construct the following two solutions.

\subsection{Solution A}

The first solution is

\begin{equation}
\Phi(t) = \frac{1}{2}\log \left(\sum_{k=1}^\infty \frac{\lambda_k}{1+\tau^{2k}}  \right), \qquad \tau\equiv  \frac{t}{t_0} = \frac{\sqrt{2d}}{\sqrt{\alpha'}}t.
\label{eq:ansatz A}
\end{equation}

\noindent The   solution (\ref{eq:2nd order solution}) is a special case with  $\lambda_1 =1/\beta^2$ and  $\lambda_{k\ge 2}=0$.
From   eqs. (\ref{eq:EOM in dilaton}), we get

\begin{eqnarray}
H(t)& =& \frac{-\left(\sum_{k=1}^{\infty } \frac{2 k \lambda _k \tau^{2 k-1}} {\left(\tau^{2 k}+1\right)^2}\right)^2 + \left(\sum _{k=1}^{\infty } \frac{\lambda _k}{\tau^{2 k}+1}\right) \sum _{k=1}^{\infty } \left(\frac{8 k^2 \lambda _k \tau^{4 k-2}}{\left(\tau^{2 k}+1\right)^3}- \frac{2 k (2 k-1) \lambda _k \tau^{2 k-2}}{\left(\tau^{2 k}+1\right)^2}\right)}{\sqrt{2} \sqrt{\alpha' } \beta  \left(\sum _{k=1}^{\infty } \frac{\lambda _k}{\tau^{2 k}+1}\right)^{5/2}}, \nonumber\\
f(H(t)) &=& - \frac{2 \sqrt{2} \beta  d}{\sqrt{\alpha'}} \sqrt{\sum _{k=1}^{\infty } \frac{\lambda _k}{\tau ^{2 k}+1}}, \nonumber\\
g(H(t)) &=& -\frac{2 d \left(\sum_{k=1}^{\infty} \frac{ k \lambda _k \tau ^{2 k-1}}{\left(\tau ^{2 k}+1\right)^2}\right)^2}{ \alpha'  \left(\sum_{k=1}^{\infty} \frac{\lambda _k}{\tau ^{2 k}+1}\right)^2}
\label{eq:general H}
\end{eqnarray}

\noindent One of the big advantages of the ansatz (\ref{eq:ansatz A}) is that as long as $\Phi(t)$ is non-singular, $H(t)$ is guaranteed to be non-singular. We therefore only need to  care about the singularity of $\Phi(t)$. Another advantage of the ansatz (\ref{eq:ansatz A}) is that every individual term inside log is non-singular, in contrast to the perturbative solution where all terms are singular. Singularities appear if and only if
\begin{equation}
\sum_{k=1}^\infty \frac{\lambda_k}{1+\tau^{2k}} = 0,
\label{eq:singular condition A}
\end{equation}
has real roots. In the perturbative regime   $t\to\infty$ ($\alpha'\to0$),  the ansatz   $\Phi(t)$  is expanded as,
\begin{eqnarray}
\Phi(t/\sqrt{\alpha'}\to\infty) &=& \frac{1}{2}\log \left(\frac{\lambda_1}{\tau^2} \right) + \frac{1}{2}\log \left(   \sum_{k=1}^\infty \frac{1}{\tau^{2k-2}}\frac{\lambda_k/\lambda_1}{1+1/\tau^{2k}}  \right)\nonumber\\
&=& \frac{1}{2}\log \left(\frac{\lambda_1}{\tau^2} \right) + \frac{1}{2}\log \left( \frac{1}{1+1/\tau^2} +   \sum_{k=2}^\infty \frac{1}{\tau^{2k-2}}\frac{\lambda_k/\lambda_1}{1+1/\tau^{2k}}  \right)\nonumber\\
&=& -\frac{1}{2}\log\left(\frac{\tau^2}{\lambda_1} \right) + \frac{\lambda _2-\lambda _1}{2 \lambda _1}\frac{1}{\tau^2}+ \frac{\lambda _1^2+2 \left(\lambda _2+\lambda _3\right) \lambda _1-\lambda _2^2}{4 \lambda _1^2}\frac{1}{\tau^4}\nonumber\\
&&- \frac{\lambda _1^3+3 \left(\lambda _2-\lambda _3-\lambda _4\right) \lambda _1^2+3
\lambda _2 \left(\lambda _2+\lambda _3\right) \lambda _1-\lambda _2^3}{6 \lambda _1^3 }\frac{1}{\tau^6} +\cdots.
\end{eqnarray}
To match the   perturbative solution (\ref{eq:perturbative solution}), the coefficients $\lambda_i$ are fixed:

%

\begin{eqnarray}
\lambda_1 &=&\frac{1}{\beta^2},\quad \lambda_2 =0, \quad  \lambda_3=\frac{4+512 c_3}{3\beta^2}, \quad \lambda_4= \frac{-4}{15\beta^2 }(31+6272 c_3 + 3072 c_4),\nonumber\\
\lambda_{5} & = & \frac{8\left(1638400c_{3}^{2}+66688c_{3}+53248c_{4}+20480c_{5}+219\right)}{35\beta^{2}},\nonumber\\
&\cdots &,
\label{eq:lambda fixing}
\end{eqnarray}
where we used $c_2=1/64$. It is clear   that  $\lambda_n$ is fixed by $c_{k\le n}$. Guaranteed by the EOM, in terms of $H(t)$, $f(H)$ and $g(H)$ in (\ref{eq:general H}) must be identical to eq. (\ref{eq:EOM fh gh}) after replacing $\lambda_n$ by   $c_{k\le n}$. And it is easy to understand that matching $H(t)$ produces the same $\lambda_i$.   The solution (\ref{eq:ansatz A}) is non-perturbative in the sense that it is defined in the whole regime $t\in (-\infty,\infty)$ and $\alpha'$ does not need to approach zero.   What we really show is that $\alpha'$ corrections do admit non-singular evolutions. Up to any order $n$, though $\lambda_{k\le n}$ are fixed by the (in the future) known $c_{k\le n}$,  one always has  freedom to choose $\lambda_{k>n}$ as any real value to violate the singular condition (\ref{eq:singular condition A}).

\subsection{Solution B}

Suppose the coefficients $c_{k\le n}$ are known, another interesting ansatz is

\begin{equation}
\Phi(t)  =  -\frac{1}{2N}\log\left[\stackrel[k=0]{N}{\sum}\rho_{k}\tau^{2k}\right], \qquad \tau\equiv  \frac{t}{t_0} = \frac{\sqrt{2d}}{\sqrt{\alpha'}}t,
\label{eq:ansatz B}
\end{equation}
\noindent where $N\ge n-1$ is some arbitrary integer and $\rho_0>0$, $\rho_N>0$.  Also from  eqs. (\ref{eq:EOM in dilaton}), we have

\begin{eqnarray}
H(t)&=& \frac{-\left(\sum _{k=0}^N \rho _k \tau ^{2 k}\right) \sum _{k=0}^N 2 k (2 k-1) \rho _k \tau ^{2 k-2} + \left(\sum _{k=0}^N 2 k \rho _k \tau ^{2 k-1}\right){}^2}{\sqrt{2} \sqrt{\alpha' } \beta  N   \left(\sum _{k=0}^N \rho _k \tau ^{2 k}\right){}^{2-\frac{1}{2 N}} },\nonumber\\
f(H(t)) &=& - \frac{2\sqrt 2 \beta d }{\sqrt{\alpha'}} \left(\sum _{k=0}^N \rho _k \tau ^{2 k}\right)^{-\frac{1}{2 N}},\nonumber\\
g(H(t)) &=&  -\frac{d \left(\sum _{k=0}^N 2 k \rho _k \tau ^{2 k-1}\right)^2}{2 \alpha'  N^2 \left(\sum _{k=0}^N \rho _k \tau ^{2 k}\right)^2}.
\label{eq:general H B}
\end{eqnarray}
The   solution (\ref{eq:2nd order solution}) is a special case with $N=1$ and $\rho_0= \rho_1 =\beta^2$. This solution has the same advantages as solution A: every single term inside log is non-singular; $\Phi(t)$ and $H(t)$ share the same singularity if and only if

\begin{equation}
\stackrel[k=0]{N}{\sum}\rho_{k}\tau^{2k}=0,
\label{eq:singular condition B}
\end{equation}
has real roots. In the perturbative regime   $t\to\infty$ ($\alpha'\to0$),  the ansatz  $\Phi(t)$ in (\ref{eq:ansatz B}) is expanded as,

\begin{eqnarray}
\Phi(t)  &=&  -\frac{1}{2} \log(\tau^2 \rho_N^{1/N}) -\frac{1}{2N} \Bigg\{ \frac{\rho_{N-1}}{\rho_N}\frac{1}{\tau^2} + \frac{2\rho_N \rho_{N-2} -\rho^2_{N-1}}{2\rho_N^2}  \frac{1}{\tau^4} +\frac{3\rho_N^2 \rho_{N-3} - 3 \rho_N \rho_{N-1} \rho_{N-2} +\rho_{N-1}^3}{3\rho_N^3} \frac{1}{\tau^6}\nonumber \\
& & +  \frac{4\rho_N^3\rho_{N-4}- 2\rho_{N}^2 \rho_{N-2}^2  -4 \rho_N^2 \rho_{N-1}\rho_{N-3} +4\rho_N \rho_{N-1}^2\rho_{N-2} -\rho_{N-1}^4 }{4\rho^4_N}\frac{1}{\tau^8} + {\cal O}\left(\frac{1}{\tau^{10}} \right) \Bigg\}.
\end{eqnarray}
To match the   perturbative solution (\ref{eq:perturbative solution}), the coefficients $\rho_i$ are fixed:
\begin{eqnarray}
\rho_N &=& \beta^{2N},\quad\rho_{N-1}=N\beta^{2N},\quad\rho_{N-2}=\frac{N\beta^{2N}}{6}\left(3N-1024c_{3}-11\right),\quad\cdots,
\end{eqnarray}
where we used $c_2 =1/64$.   It should be noted that  only $\rho_N, \rho_{N-1} \cdots \rho_{N-n+1}$ are fixed by the known coefficients $c_1,c_2\cdots c_n$. Other parameters $\rho_0, \rho_1 \cdots \rho_{N-n}$ can take any real numbers to violate the singular condition (\ref{eq:singular condition B}). In particular, we should set $\rho_0 >0$ to avoid $t=0$ becoming a singularity.
Again, guaranteed by the EOM, in terms of $H(t)$, $f(H)$ and $g(H)$ in (\ref{eq:general H B}) must be identical to eq. (\ref{eq:EOM fh gh}) after replacing $\rho_{k\ge N-n+1}$ by   $c_{k\le n}$.

We close this section by rewriting the EOM (\ref{eq:Original EoM}) in another form, which will be used to derive a (phenomenological) map between $\alpha'$ corrected EOM and loop corrected EOM in next section. Note from eq. (\ref{eq:gf relation}), we have
\begin{equation}
\dot g (H) = g'(H) \dot H(t) = H f'(H) \dot H(t) = H\dot f(H),
\end{equation}
Then one can easily verify that the  EOM (\ref{eq:Original EoM}) can be recast as
\begin{eqnarray}
2\ddot{\Phi}-2 d f(H)^2 +\frac{d}{dt}\Big[g(H)+  d f(H)^2    \Big]\, \frac{f(H)}{\dot f(H)}  &=&0,\nonumber \\
\dot{\Phi}^{2}-df\left(H\right)^{2}+\left[g\left(H\right)+df\left(H\right)^{2}\right] & = & 0,\nonumber \\
\dot{f}\left(H\right)-f\left(H\right)\dot{\Phi} & = & 0.\label{eq:re EOM}
\end{eqnarray}

%
%

%
%
%
%

\section{A  map between $\alpha'$ corrected EOM and loop corrected EOM  }

It was discovered long time ago that the big-bang singularity could be regularized by loop corrections. Referring to the complete string effective action (\ref{eq:complete action}), setting $\alpha'=0$, we are left with a purely loop corrected theory. All higher genus corrections have the same structure as the tree level, but with unknown coefficients and different couplings. In the context of discussing singularity resolution,   it is sufficient to implement some effective dilaton potentials to stand for loop corrections. However, since the physical dilaton $\phi$ is not an $O(d,d)$ scalar, a generalized non-local dilaton is introduced to keep the  $O(d,d)$ symmetry \cite{Gasperini:1992em,Gasperini:2003pb},

\begin{equation}
e^{-\Phi\left(x\right)}=\int d^{d+1}x^{\prime}\sqrt{-g\left(x^{\prime}\right)}e^{-2\phi\left(x^{\prime}\right)}\sqrt{4| g^{\mu\nu}\partial_{\mu}\phi\left(x^{\prime}\right)\partial_{\nu}\phi\left(x^{\prime}\right)|} \delta\left(2\phi\left(x^{\prime}\right)-2\phi\left(x\right)\right),
\end{equation}
which reduces to the $O(d,d)$ dilaton in the FLRW background (\ref{FLRW}),
\begin{equation}
e^{-\Phi\left(t\right)}=V_{d}\int dt'\left|\frac{d\left(2\phi\right)}{dt'} \right|\sqrt{-g\left(t'\right)}e^{-2\phi\left(t'\right)}\delta\left(2\phi\left(t\right)-2\phi\left(t^{\prime}\right)\right)=V_{d}\sqrt{-g\left(t\right)}e^{-2\phi\left(t\right)}.
\label{eq:reduced non-local dilaton}
\end{equation}
A phenomenological loop corrected effective theory then is
\begin{eqnarray}
I_{\rm Loop}&=&\int d^{d+1}x\sqrt{-g}e^{-2\phi}\left[R+4\left(\partial_{\mu}\phi\right)^{2} - V\left(e^{-\Phi\left(x\right)}\right)\right],\nonumber\\
&=& \int dt e^{-\Phi} \big[ -\dot\Phi + d H^2 -V(e^{-\Phi})\big],
\label{eq:action with potential}
\end{eqnarray}
where in the second line, we applied the FLRW background. The EOM is \cite{Gasperini:1992em,Gasperini:2003pb},

\begin{eqnarray}
2\ddot{\Phi} -2 dH^2 - \frac{\partial V}{\partial\Phi} &=&0,\nonumber \\
\dot{\Phi}^{2}-dH^{2}-V & = & 0,\nonumber \\
\dot{H}-H\dot{\Phi} & = & 0.\label{eq:non-local cosmo EoM}
\end{eqnarray}
Using the third equation, we have
\begin{equation}
\frac{\partial V}{\partial\Phi} =\frac{dV(\Phi)}{dt} \frac{1}{\dot\Phi} = \frac{dV}{dt} \frac{H(t)}{\dot H(t)}.
\end{equation}
Therefore, the EOM (\ref{eq:non-local cosmo EoM}) can be rewritten as

\begin{eqnarray}
2\ddot{\Phi} -2 dH^2 - \frac{dV}{dt} \frac{H(t)}{\dot H(t)} &=&0,\nonumber \\
\dot{\Phi}^{2}-dH^{2}-V & = & 0,\nonumber \\
\dot{H}-H\dot{\Phi} & = & 0.\label{eq:loop EOM}
\end{eqnarray}

\noindent Comparing with the $\alpha'$ corrected EOM (\ref{eq:re EOM}), we immediately identify a map between   the loop corrected EOM and the $\alpha'$ corrected EOM,

\begin{eqnarray}
\alpha^{\prime}\;\mathrm{EOM}\;\eqref{eq:re EOM}: &  & \mathrm{Loop\;EOM}\;\mathrm{\eqref{eq:loop EOM}:}\nonumber \\
g\left(H_{\alpha'}\right)+df\left(H_{\alpha'}\right)^{2} & \longleftrightarrow & -V_L,\nonumber \\
f\left(H_{\alpha'}\right) & \longleftrightarrow & H_L,\nonumber \\
\Phi_{\alpha'} & \longleftrightarrow & \Phi_L +\Phi_0,\label{eq:identification}
\end{eqnarray}

\noindent where $\Phi_0$ is a constant and  the subscripts $L$  and $\alpha'$  indicate to what corrections the quantities belong. It should be noted that in order to match the perturbative solution, we need to rescale $f(H_{\alpha'})=-2d H_{\alpha'} +\cdots$ by dividing $-2d$ after the mapping.
This effectively can be accomplished by the constant $\Phi_0$.  We want to stress that this does not mean there must exist such a map between the \emph{true} complete loop corrections and complete $\alpha'$ corrections, since they might not share the same solution $\Phi (t)$ and the action (\ref{eq:action with potential}) is a greatly simplified model. However, this \emph{phenomenological} but instructive map is still very useful to mutually generate new solutions for either of them.

\vspace{2ex}

\noindent\underline{Generate $\alpha'$ corrected solutions from loop corrected solutions}


\noindent In \cite{Gasperini:2003pb, Gasperini:2004ss},  a  class of phenomenological loop corrected solutions was constructed,

\begin{eqnarray}
\Phi^{(n)}_L(t) & = & \frac{1}{2n}\log\left(\frac{\sigma_n^{2n}}{1+\left(m_n t\right)^{2n}}\right),\nonumber\\
H^{(n)}_{L} (t) & = & \frac{1}{\sqrt d} \frac{m_n}{\sigma_n} e^{\Phi^{(n)}_L \left(t\right)}= \frac{m_n}{\sqrt d} \left[\frac{1}{1+\left(m_n t\right)^{2n}}\right]^{1/2n}.
\label{eq:nth Phi}
\end{eqnarray}
with a potential
\begin{equation}
V^{(n)}_{L}  =   \left(\frac{m_n}{\sigma_n}\right)^{2}e^{2\Phi^{(n)}_L \left(t\right)}\left[\left(1 - \sigma_n^{-2n} e^{2n\Phi^{(n)}_L \left(t\right)}\right)^{\frac{2n-1}{n}}-1\right].
\label{eq:potentials}
\end{equation}
where $n$ is any positive integer and $\sigma_n$ is a dimensionless coefficient. Since $e^\Phi$ roughly plays the role of a ``dimensionally reduced'' coupling constant, the parameter $n$ is effectively  a  ``loop counting'' parameter and the potential (\ref{eq:potentials}) could be interpreted as the non-perturbative contributions from $n$th loop.

Using the identification (\ref{eq:identification}), it is straightforward to generate a class of  $\alpha'$ corrected solutions.  By matching the perturbative $\alpha'$ corrected solution (\ref{eq:perturbative solution}),  the parameters $n$, $\sigma_n$ and $m_n$ are fixed

\begin{equation}
n=1,\quad \sigma_1 =\frac{1}{\beta},\quad m_1 = \sqrt\frac{2d}{\alpha'}.
\label{eq:n choice}
\end{equation}

\noindent Happily, the generated  $\alpha'$ corrected solution is nothing but the solution (\ref{eq:2nd order solution}), which was constructed in \cite{Wang:2019kez}.

\vspace{2ex}

\noindent\underline{Generate loop corrected solutions from $\alpha'$ corrected solutions A}


\noindent  In section $2$, we constructed a general class of $\alpha'$ corrected solutions (\ref{eq:ansatz A}) and (\ref{eq:general H}), with the parameters $\lambda_i$ fixed by matching the perturbative solution as in eq. (\ref{eq:lambda fixing}). Applying the identification (\ref{eq:identification}), we obtain a general class of loop corrected solutions


\begin{eqnarray}
\Phi_L(t) &=&  \frac{1}{2}\log \left(\sum_{n=1}^\infty \frac{\sigma_n^{2n}}{1+(m_n\, t)^{2n}}  \right) = \frac{1}{2} \log\left( \sum_{n=1}^\infty e^{2n \Phi^{(n)}_L}  \right),  \nonumber\\
H_L(t)&=& \frac{1}{\sqrt d} \frac{m_1}{\sigma_1} \sqrt{\sum_{n=1}^\infty \frac{\sigma_n^{2n}}{1+(m_n\, t)^{2n}}}= \frac{1}{\sqrt d} \frac{m_1}{\sigma_1}  \sqrt{\sum_{n=1}^\infty e^{2n \Phi_L^{(n)}}},\nonumber\\
V_L(\Phi_L^{(n)})&=& \left(\frac{\sum_{n=1}^\infty n \dot\Phi_L^{(n)} e^{2n\Phi_L^{(n)}}}   {\sum_{n=1}^\infty e^{2n \Phi^{(n)}_L} }\right)^2 -\left(\frac{m_1}{\sigma_1}\right)^2 \sum_{n=1}^\infty e^{2n \Phi^{(n)}_L},
\label{eq:loop solution}
\end{eqnarray}
where we used eq. (\ref{eq:nth Phi}) to express quantities in term of the $n$th loop contributions. Although it is straightforward to verify that $\dot\Phi_L^{(n)}$ can   be expressed in term of $e^{\Phi_L^{(n)}}$ from eq. (\ref{eq:nth Phi}), we keep   $\dot\Phi_L^{(n)}$ to leave the freedom of the constants in  $\Phi_L^{(n)}$. Thus the potential $V_L(\Phi_L^{(n)})$ is a function of all $n$th loop contributions. In practice, since $\sigma_n$'s are free constants, all $m_n$ can be set to be the same as $m_1$ without losing generality. $m_1/\sigma_1$ is going to be fixed by the tree level solution up to an integration constant. $\sigma_2$ is going to be fixed by   the  one loop correction,    $\sigma_3$ is going to be fixed by   the  two loop correction, and so on. It is difficult to find this solution   directly from the loop corrected EOM (\ref{eq:non-local cosmo EoM}). We thus generate infinitely many new solutions  for  loop corrections.

Some remarks  are in order. Let us first expand $\Phi_L^{(n)}$ in eq. (\ref{eq:nth Phi}) as $|t|\to \infty$

\begin{equation}
\Phi_L^{(n)} = -\frac{1}{2}\log \frac{t^2}{t_0^2} + {\cal O}\left( \frac{t_0^{2n}}{t^{2n}} \right), \qquad |t|\to \infty.
\end{equation}
For loop corrections, we only know the tree level results, and coefficients of higher loops are still out of reach. Therefore, when constructing loop corrected solutions, one only needs to match the tree level perturbative solution $\Phi_L = -\frac{1}{2}\log (t^2/t_0^2)+\cdots$.  This is why in eq. (\ref{eq:nth Phi}) $n$ can be any positive integer. But this is not consistent, since the loop corrections should be introduced order by order as one loop, two loop, and so on. The solution (\ref{eq:loop solution})  we constructed is   much more reasonable, since all loops are included and when higher loop corrections are given, more $\sigma_i$'s are fixed.

On the other hand, we now know much more information about the $\alpha'$ corrections. Not only the coefficient of the first order in $\alpha'$, the behaviors of the higher orders are also determined by the Hohm-Zwiebach action. These information selects (\ref{eq:n choice}), particularly $n=1$, out of other numbers.


\vspace{2ex}

\noindent\underline{Generate loop corrected solutions from $\alpha'$ corrected solutions B}

\noindent We can also generate loop corrected solutions from solution B (\ref{eq:ansatz B}) and (\ref{eq:general H B}) with the   identification (\ref{eq:identification}),


\begin{eqnarray}
\Phi_L (t)&=& -\frac{1}{2N} \log\Bigg[\sum_{n=0}^N  (m_n t)^{2n}\Bigg] = -\frac{1}{2N} \log \Big[ \sum_{n=0}^N  e^{-2n \Phi_L^{(n)}}\Bigg], \nonumber\\
H_L(t) &=& \frac{m_N }{\sqrt d} \Bigg[\sum_{n=0}^N   (m_n t)^{2n}\Bigg]^{-1/2N} = \frac{m_N }{\sqrt d} \Bigg[ \sum_{n=0}^N  e^{-2n \Phi_L^{(n)}}\Bigg]^{-1/2N}, \nonumber\\
V_L (\Phi^{(n)}) &=& \left( \frac{\sum_{n=0}^N n \dot\Phi_L^{(n)} e^{-2n\Phi_L^{(n)}}       }{N\sum_{n=0}^N   e^{-2n\Phi_L^{(n)}}}                   \right)^2 - m_N^2  \Bigg[ \sum_{n=0}^N  e^{-2n \Phi_L^{(n)}}\Bigg]^{-1/N},
\end{eqnarray}
where   we absorbed various constants into $\Phi_L^{(n)}$. Again, we used (\ref{eq:nth Phi}) to express quantities in term of the $n$th loop contributions.   $m_N$ is an integration constant. $m_{N-1}$ is going to be fixed by   the  one loop correction,    $m_{N-2}$ is going to be fixed by   the  two loop correction, and so on. It is also not easy to find this solution   directly from the loop corrected EOM (\ref{eq:non-local cosmo EoM}).

\section{Conclusion}

In this paper, we provided  two formulas  to construct $\alpha'$ corrected cosmological solutions without curvature singularities, for bosonic gravi-dilaton system. Once the coefficient  $c_{n>2}$ of the $n$th order in $\alpha'$ expansion is provided,  more accurate  solutions can be constructed straightforwardly. We can always make the solution non-singular by  adjusting $\lambda_{k>n}$ ($\rho_{k<N-n}$) freely. We also gave a phenomenological map between the $\alpha'$ corrected EOM and loop corrected EOM. Although this map is based on  considerably simplified loop corrections, one can use it to generate new solutions. Especially the loop corrected solutions generated from the $\alpha'$ corrected solutions are more reasonable than those in literature.

We addressed vacuum scenario in this work and set     $b_{\mu\nu}=0$. Since the theory is supposed to be $O(d,d)$ invariant, one might  rotate time dependent $b_{\mu\nu}(t)$ into the evolution to get some new features. For example, nontrivial $b_{\mu\nu}(t)$ could stabilize the string coupling as $t\to\infty$, as showed in \cite{Gasperini:1991qy}   for loop corrected solutions. Also matter sources in an $O(d,d)$ fashion are expected to lead to more realistic configurations.

Though looks quite phenomenological, in some sense, the map between the $\alpha'$ corrected EOM and loop corrected EOM we found is actually  suggestive. In a previous work \cite{Wang:2017mpm}, we   conjectured a possible correspondence between genus expansion and $\alpha^{\prime}$ expansion by noting that, in terms of Riemann normal coordinate,  the $\alpha'$ expansion of a string propagating in AdS     matches exactly        the genus expansion in the Goparkumar-Vafa formula, order by order. So there might exist some deep connection between $\alpha^{\prime}$ expansion and loop expansion. To gain more insight, we need more information about the loop expansion.

In this paper, we  used the map to construct new solutions. It is reasonable to expect there are more applications, at least phenomenologically. In coming works,  we will address some inspiring applications.

\vspace{5mm}

\noindent {\bf Acknowledgements}
This work is supported in part by the NSFC (Grant No. 11875196, 11375121, 11005016 and 11947225).

\end{document}